\documentclass[pdftex,twocolumn,epjc3]{svjour3}          

\RequirePackage[T1]{fontenc}

\smartqed  

\RequirePackage{graphicx}
\RequirePackage{multirow}
\RequirePackage{booktabs}
\RequirePackage{threeparttable}
\RequirePackage{amsmath}
\RequirePackage{mathptmx}      
\RequirePackage{flushend}
\RequirePackage[numbers,sort&compress]{natbib}
\RequirePackage[colorlinks,citecolor=blue,urlcolor=blue,linkcolor=blue]{hyperref}

\journalname{Eur. Phys. J. C}

\begin{document}

\title{Study on cosmogenic activation in copper for rare event search experiments}

\author{Ze She\thanksref{addr1} \and Zhi Zeng\thanksref{addr1} \and Hao Ma\thanksref{e1,addr1} \and 
Qian Yue\thanksref{addr1} \and Mingkun Jing\thanksref{addr1} \and Jianping Cheng\thanksref{addr1,addr2} \and
Junli Li\thanksref{addr1} \and Hui Zhang\thanksref{addr1}
}

\thankstext{e1}{mahao@mail.tsinghua.edu.cn}

\institute{Key Laboratory of Particle and Radiation Imaging (Ministry of Education) and Department of Engineering Physics, Tsinghua University, Beijing 100084, China \label{addr1} 
 	   \and
	   College of Nuclear Science and Technology, Beijing Normal University, Beijing 100875, China \label{addr2}
}

\date{Received: date / Accepted: date}

\maketitle

\begin{abstract}
The rare event search experiments using germanium detectors are performed in the underground laboratories to prevent cosmic rays. 
However, the cosmogenic activation of the cupreous detector components on the ground will generate long half-life radioisotopes and contribute continually to the expected background level. 
We present a study on the cosmogenic activation of copper after 504 days of exposure at an altitude of 2469.4 m outside the China Jinping Underground Laboratory (CJPL).
The specific activities of the cosmogenic nuclides produced in the copper bricks were measured using a low background germanium gamma-ray spectrometer at CJPL. 
The production rates at sea level, in units of nuclei/kg/day, are ${18.6 \pm 2.0}$ for ${^{54}}$Mn, ${9.9 \pm 1.3}$ for ${^{56}}$Co, ${48.3 \pm 5.5}$ for ${^{57}}$Co, ${51.8 \pm 2.5}$ for ${^{58}}$Co and ${39.7 \pm 5.7}$ for ${^{60}}$Co, respectively. 
Given the expected exposure history of the germanium detectors, a Monte Carlo simulation is conducted to assess the cosmogenic background contributions of the detectors' cupreous components. 
\end{abstract}

\section{Introduction}
\label{intro}
The rare event search experiments, such as dark matter direct detections and neutrinoless double beta decay experiments, are operated in deep underground laboratories with passive and active shields as well as selecting radiopure materials to effectively reduce their intrinsic background \cite{Heusser1995, Abgrall2016, Akerib2020, Ma2021}.
Under the ultra-low background achieved by substantial efforts, the radioactive impurities in the selected materials induced by cosmogenic activation on the ground could become even more prominent than the residual contamination of the primordial nuclides. 
\par
With high radiopurity and attractive mechanical properties, copper is widely innermost used as shields or parts of the detectors in rare event search experiments \cite{Jiang2018,Agostini2017,Abgrall2014}. 
However, radioisotopes produced by the cosmogenic activation during manufacture, transport, and storage, disrupt the background suppression in copper.
Among these cosmogenic nuclides, short-lived nuclides like ${^{59}}$Fe and ${^{48}}$V decay rapidly when stored in the underground laboratories.
In contrast, long-lived nuclides like ${^{60}}$Co and ${^{54}}$Mn remain and continue to contribute to the detector's background level \cite{Mei2009, Ma2018}. 
\par
The China dark matter experiment (CDEX) aiming at direct dark matter detection and neutrinoless double beta decay of ${^{76}}$Ge operates p-type point-contact germanium detectors (PPCGe) at the China Jinping underground Laboratory (CJPL) \cite{Jiang2018, Liu2019, Yang2019, She2020, Wang2020}. 
The copper used in the PPCGe detectors of the CDEX experiment leads to a crucial background contribution.
It is necessary to assess the background due to cosmogenic activations in copper to establish the background model for the CDEX experiment.
However, the production rates of cosmogenic activations differs significantly among different simulation packages according to previous reseraches \cite{Cebri2010, Wei2017}
\par
In this study, we present the measurements of the cosmogenic activation in the copper samples before and after exposure to the cosmic rays. 
In comparison with the measured results, the corresponding radionuclide production rates are also calculated by Monte Carlo simulations.
Finally, we simulate the background spectra of cosmogenic radionuclides in the copper components of the PPCGe detector array used in the future CDEX experiment under the expected exposure history of detectors above the ground.
\par

\section{Experiment and methods} 
\begin{figure}[htb]
    \centering
	\includegraphics[width=\columnwidth]{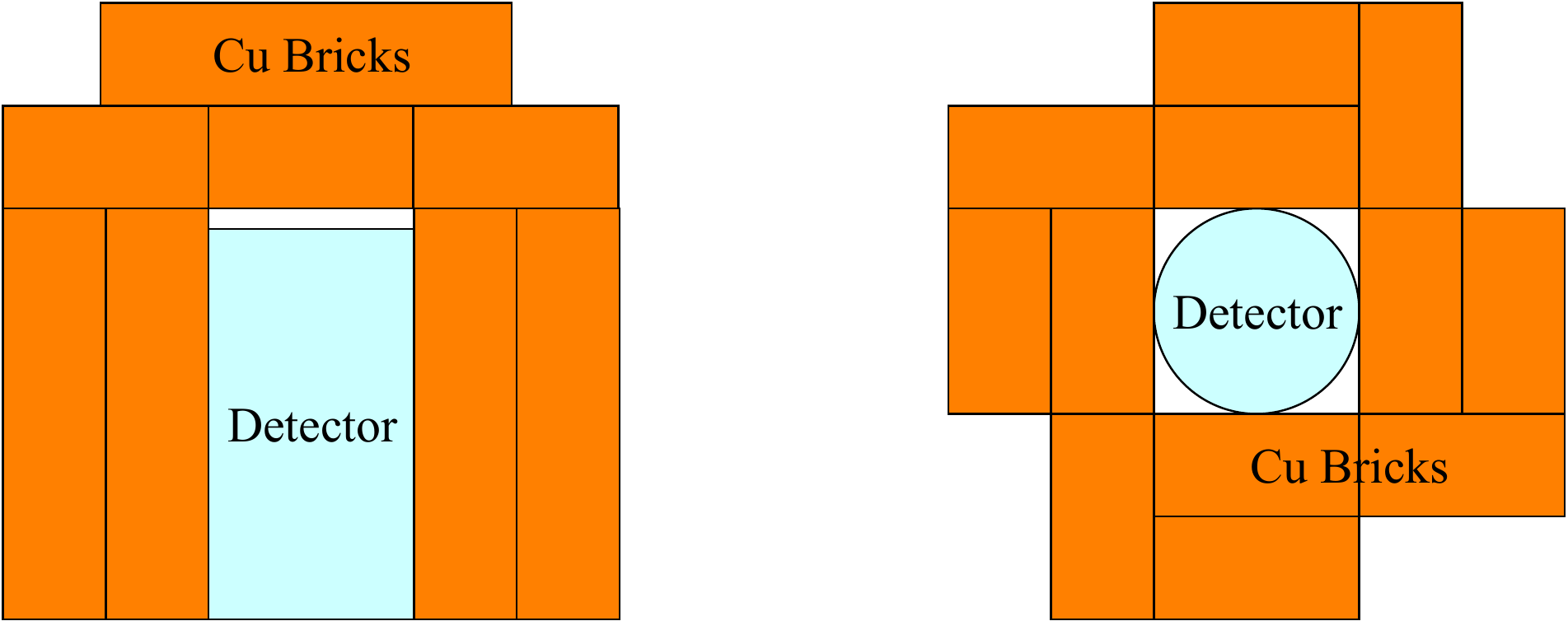}
    \caption{\textrm{ The schematic diagram of the palcement of the OFHC copper bricks in the spectrometer: right view (left) and top view (right).}}
    \label{shielding}
\end{figure}

\begin{figure}[htb]
    \centering
    \includegraphics[width=\columnwidth]{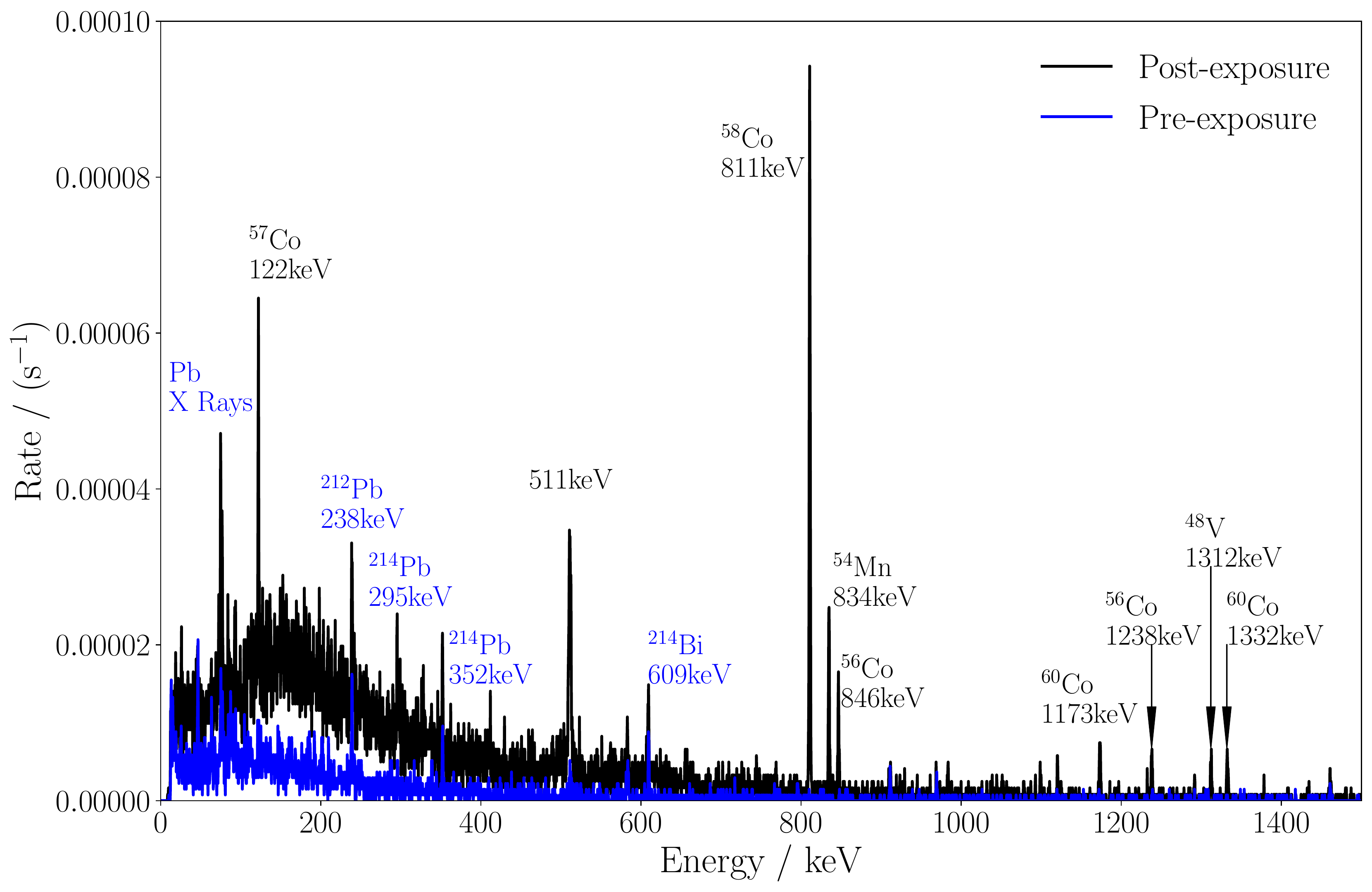}
    \caption{The measured spectra of the OFHC copper samples pre- (in blue) and post-exposure (in black). The gamma-ray peaks induced by cosmogenic activations and primordal radionuclides are also indicated. }
    \label{exposure}
\end{figure}

The oxygen-free high thermal conductivity (OFHC) copper bricks (their chemical purities are more than 99.995\% with natural isotope abundance), 20 ${\times}$ 10 ${\times}$ 5 cm in size, were exposed to cosmic rays and measured in this study.
The cosmic-ray muon flux in the CJPL is suppressed by 8 orders of magnitude compared to surface laboratories due to the 2400 m rock overburden \cite{Wu2013, Guo2021}.
These OFHC copper bricks have been housed in the CJPL for more than 4 years. 
These OFHC copper samples were measured to evaluate the residual cosmogenic radioactivity in advance with a low background germanium gamma-ray spectrometer GeTHU at CJPL, whose background rates between 100 and 2700 keV is 732 counts/day \cite{Zeng2014}. 
To enhance the detection efficiency, the germanium detector of GeTHU were surrounded by these copper sample bricks as shown in Figure~\ref{shielding}.
The measured spectrum is depicted in Figure~\ref{exposure} and the characteristic gamma-ray peaks from cosmogenic radionuclides are not statistically significant and lower than the minimum detectable activities (MDA) of GeTHU.
\par
Sixteen copper bricks with a total mass of 142.50 kg were exposed to cosmic-rays for 504 days, from Aug. 31, 2019 to Jan. 16, 2021, at an altitude of 2469.4 m in the vinicity of CJPL. 
The specific activities of cosmogenic radionuclides inside these copper bricks are calculated by spectral analysis.  
In spectral analysis, the net area of a peak is determined under the assumption of a linear background \cite{Gilmore2008}. 
Besides, the detection efficiencies for concerned radionuclides are simulated by Geant4.10.06 \cite{Agostinelli2003, Allison2006, Allison2016}. 
\par 
The specific activities of cosmogenic radionuclides can be deduced by combining the production rates and the decay rates as shown in Figure~\ref{yield}. 
\par
\begin{equation}
\label{yield}
  A = R(1-e^{-\lambda t})+A_0e^{-\lambda t},
\end{equation}
Where ${A}$ is the specific activity in the unit of Bq/kg and ${R}$ denotes the production rate of a particular cosmogenic radionuclide in unit of Bq/kg. 
${\lambda}$ represents the decay constant and ${t}$ is the time during cosmogenic activation. 
${A_0}$ stands for specific activities before exposure. 
The specific activities of the cosmogenic radionuclides are unsaturated after 504 days of exposure for radionuclides with long half-lives, such as ${^{60}}$Co. 
We have applied the exponential hypothesis as shown in (\ref{yield}) with zero ${A_0}$ to calculate the production rates due to the copper sample measurement mentioned above. 
The decay of cosmogenic radionuclides during the measurements was also considered, especially for those with short half-lives, such as ${^{58}}$Co.
\par
The cosmogenic production rate ${R}$ can be calculated and simplified as shown in (\ref{activation}).
\begin{equation}
    \label{activation}
    R = \int \frac{d\Phi(E)}{dE}\sigma(E) dE \approx \Phi_{tot}\int \frac{d\phi(E)}{dE}\sigma(E) dE \propto \Phi_{tot},
\end{equation}
${\frac{d\Phi(E)}{dE}}$ is the energy spectrum of cosmic rays while the ${\frac{d\phi(E)}{dE}}$ is the normalized energy spectrum.
${\Phi_{tot}}$ is the flux of cosmic rays.
As the energy spectra differ slightly at different altitudes within 20 km \cite{Goldhagen2002}, the normalized energy spectra at different altitudes can be simplified as a constant.
The
efore, the cosmogenic production rates are directly proportional to the strength of the cosmic-ray flux. 
The cosmic-ray flux at a given altitude can be estimated by exponential correction \cite{Lifton2014}.
The altitude corrections on the cosmic-ray flux according to (\ref{Intensity}) is applied to calculate the cosmic-ray flux at different altitudes.
\begin{equation}
    \label{Intensity}
    \Phi_{tot, H} = \Phi_{tot, 0}e^{\frac{p(H)-p(0)}{L}},
\end{equation}
where ${\Phi_{tot, H}}$ and ${\Phi_{tot, 0}}$ indicate the flux of certain particles in the cosmic ray at altitudes ${H}$ and sea level in unit of particles/cm${^2}$/s, respectively. 
${p(H)}$ is the altitude pressure at altitude ${H}$ \cite{Ziegler1996, Lifton2014}and ${L}$ is the typical absorption length and specified for a certain cosmic-ray component.
It is convinced that the production rates of the cosmogenic activations are dominated by neutrons so that the absorption length of the neutron is a suitable choice \cite{Breier2020, Cebri2020}. 
The cosmogenic production rates are proptional to the neutron flux leading to the same exponential correction can be applied to get the corresponding production rates at sea level as shown in (\ref{production})
\begin{equation}
    \label{production}
    R(H) = R(0)e^{\frac{p(H)-p(0)}{L}},
\end{equation}
For comparison with measurements, the cosmogenic production rates are also simulated by Geant4 with the physical list QGSP\_INCLXX \cite{Allison2006,Allison2016,Apostolakis2009}. 
CRY-1.7 \cite{Hagmann2007} provides the energy spectra of different cosmic-ray components (neutron, proton, muon, gamma, electron and pion) at sea level for the Geant4 simulation. 
In the simulation, we count each product after the interactions with the copper sample for the production rate calculation. 

\section{Cosmic ray activation in copper}
\label{Activation}
These copper bricks were measured from Jan 16th to 30th, 2021 and the spectrum after exposure is also shown in Figure~\ref{exposure}, in which the selected gamma-ray peaks of cosmogenic radionuclides are marked.
In addition to the characteristic gamma-ray peaks from cosmogenic radionuclides, there also exist several gamma-ray peaks related to the primordial radionuclides and the 511 keV annihilation peak.
Compared with pre-expoure spectrum, the background level gets worse due to the break-down of the low-radon air system of the laboratory when measuring these bricks after exposure.
GeTHU has a 
efilling system flushing its sample space with the low-radon air (with boil-off nitrogen gas previously) to mitigate radon and its daughters \cite{Zeng2014}.
Without flushing, the concentrations of radon and its progenies in the sample space get higher and contribute to the excessive background rates and higher gamma-ray peak rates at a lower energy region (from 238 keV to 609 keV) as shown in the Figure~\ref{exposure}.
However, it is evident that these cosmogenic gamma-ray peaks are still significant enough to calculate their specific activities.
Following the aforementioned gamma-ray spectral analysis method, the specific activities of these cosmogenic radionuclides are calculated and listed in Table~\ref{nuclides}. 
\par
\begin{table}
    \centering
    \caption{The measured specific activities of cosmogenic radionuclides after exposure (with 1 ${\sigma}$ uncertainty) in copper samples. The gamma-ray peaks with better statistics are selected to determine the specific activities when radionuclides have two or more gamma-ray peaks.}
    \label{nuclides}
    \begin{threeparttable}
        \begin{tabular*}{\columnwidth}{@{\extracolsep{\fill}}llll@{}}
        \toprule
        Radionuclides& Half-lives / (d)& Specific activity / (mBq/kg)\\ 
        \hline
            $^{54}$Mn&  312.20&  ${0.871 \pm 0.095}$\\
            $^{56}$Co&   77.24&  ${0.649 \pm 0.084}$\\
            $^{57}$Co&  271.74&  ${2.423 \pm 0.275}$\\
            $^{58}$Co&   70.86&  ${3.394 \pm 0.166}$\\
            $^{60}$Co& 1925.28&  ${0.464 \pm 0.070}$\\ 
        \bottomrule
        \end{tabular*}
    \end{threeparttable}
\end{table}
After the altitude correction, the cosmogenic production rates are reported in \c
ef{research}, compared with results from previous studies.
The results of our measurements are generally consistent with those by the simulation with Geant4 except for ${^{56}}$Co. 
The specific activity of ${^{57}}$Co is about 3\% smaller than simulation while that of ${^{56}}$Co is 49\% of the simulation result.
In addition, Geant4 simulation result in \cite{Zhang2016} is only half of our simulated result for ${^{56}}$Co mainly due to the differences of the physical lists between 'shielding' and 'QGSP\_INCLXX'and copper samples used in these two simulations.
Due to its short half-life, 77.2 days, the ${^{56}}$Co will decay rapidly when stored underground resulting in a negiligible background contribution.
Thus, the difference between simualtion and measurement is not the prominent obstacle to assess the cosmogenic background of the underground experiments.  
Our measurements are relatively close to those measured by Ref. \cite{Baudis2015}.
The discrepancies of the production rates in the literatures could be due to the different cross-section databases, uncertainties of altitude correction, and the fluxes and energy spectra of cosmic rays in different experimental latitudes.
\par
\begin{table*}[htb]
    \begin{threeparttable}
        \centering
        \caption{Production rates of cosmogenic radionuclides in the copper bricks at sea level (Unit: kg${^{-1}}$ d${^{-1}}$).}
        \label{research}
        \begin{tabular*}{\textwidth}{@{\extracolsep{\fill}}lrrrrrl@{}}
            \toprule
            & \multirow{2}*{Method}& \multicolumn{5}{c}{Nuclides} \\
            \cmidrule{3-7} & & ${^{54}}$Mn& ${^{56}}$Co& ${^{57}}$Co& ${^{58}}$Co& ${^{60}}$Co\\
            \hline
            \multirow{2}*{This work}& Measurement& ${18.6 \pm 2.0}$& ${9.9 \pm 1.3}$& ${48.3 \pm 5.5}$&	${51.8 \pm 2.5}$& ${39.7 \pm 5.7}$\\
            ~& Geant4& 21.1& 20.4& 49.6& 70.9& 44.1\\[5pt]
            R. Breier et al \cite{Breier2020}& CONUS \cite{Masarik1999}& 14& 10& 50& 76& 92\\[5pt]
            \multirow{2}*{C. Zhang et al\cite{Zhang2016}}& Geant4& 12.31& 10.32& 67.15& 57.26& 64.63\\
            ~& ACTIVIA\tnote{1}& 14.32/30.00& 8.74/20.13& 32.44/77.45& 56.61/138.06& 26.28/66.12\\[5pt]
            S. Cebrian et al \cite{Cebri2010}& MENDL+YIELDX\tnote{2}& 32.5/27.7& 22.9/20.0& 88.3/74.1& 159.6/123.0& 97.4/55.4\\[5pt]
            \multirow{2}*{L. Baudis et al\cite{Baudis2015}}& Measurement& ${13.3^{+3.0}_{-2.9}}$& ${9.3^{+1.2}_{-1.4}}$& ${44.8^{+8.6}_{-8.2}}$& ${68.9^{+5.4}_{-5.0}}$& ${29.4^{+7.1}_{-5.9}}$\\
            ~& COSMO & 13.5& 7.0& 30.2& 54.6&25.7\\[5pt]
            M. Laubenstein et al\cite{Laubenstein2009}& Measurement& ${8.85\pm0.86}$& ${9.5\pm1.2}$& ${73.8\pm16.7}$& ${67.9\pm3.7}$& ${86.2\pm7.6}$\\[5pt]
        \bottomrule
        \end{tabular*}
        \begin{tablenotes}
            \item[1]{These two calculations rely on two cosmic neutron spectra from Gordan and ACTIVIA \cite{Gordan2004}}
            \item[2]{These two calculations rely on two cosmic neutron spectra from Ziegler and Gordan, respectively \cite{Gordan2004,Ziegler1998}}
    \end{tablenotes}
    \end{threeparttable}
\end{table*}
\par

\section{Background assessment in PPCGe detector}
The next phase of CDEX experiment, named CDEX-100, will operate about 100 kg germanium detectors immersed in the liquid nitrogen in a 1725-m$^{3}$ cryotank at the Hall-C of the second phase of CJPL.
The specific activities of cosmogenic radionuclides are calculated based on their production rates and the expected exposure hitory. 
The total exposure period is a combination of manufacture, transportation above ground and storage underground.
Their specific activities can be calculated as (\ref{activity})
\begin{align}
    A = & R(0)e^{p(H_1)/L}(1-e^{-\lambda t_{manu}})e^{-\lambda(t_{trans}+t_{cool})} \nonumber \\ 
	&+ F \times R(0)e^{p(H_2)/L}(1-e^{-\lambda t_{trans}})e^{-\lambda t_{cool}}, 
    \label{activity}
\end{align}
The ${R(0)}$ is the measured production rates at sea level shown in the Table~\ref{research}.
The ${H_1}$ and ${H_2}$ are the altitudes of the manufacture factory and the transportation routine above ground.
The time of manufacture, transportation above ground and storage underground are represented with ${t_{manu}}$, ${t_{trans}}$ and ${t_{cool}}$, respectively.
The factor ${F}$ is the factor describing the shielding effect when transportation with a steel or lead surroudings. 
\par
\begin{figure}[htb]
    \centering
    \includegraphics[width=\columnwidth]{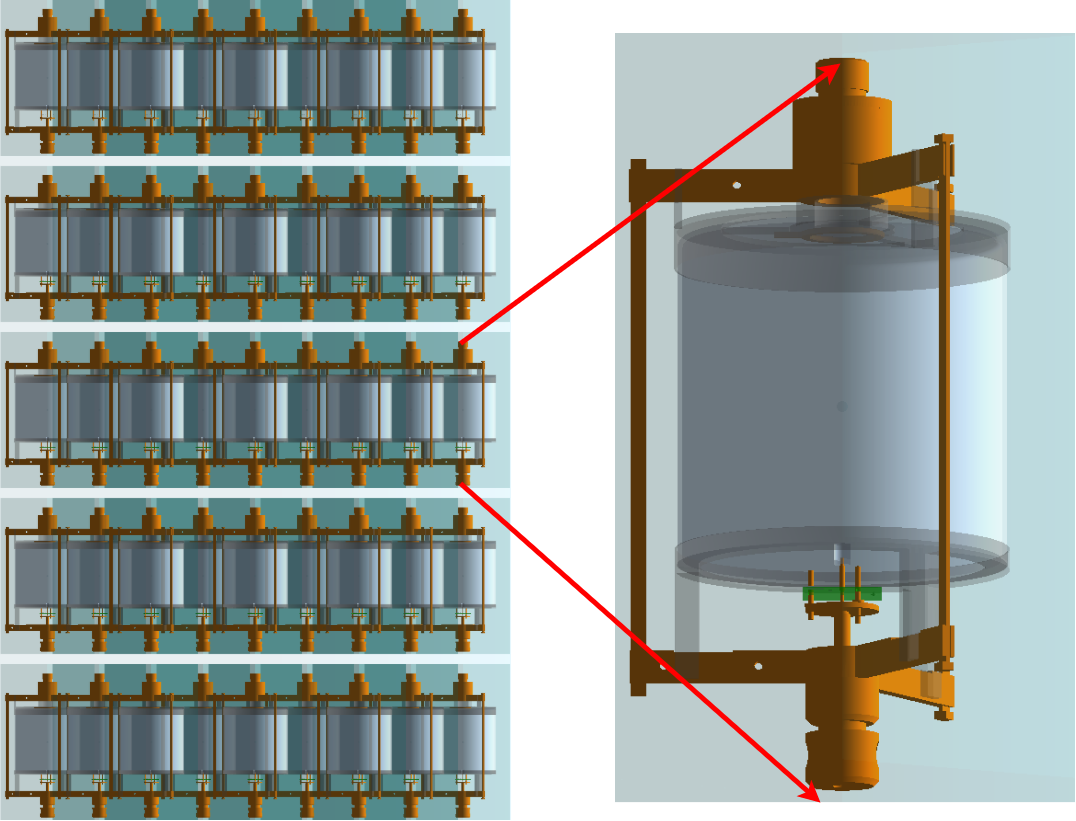}
    \caption{The sketch of the PPCGe detector array (left) and the PPCGe detector unit (right) implemented in SAGE together with its cupreous components shown in orange. The PPCGe detector array is immersed in the liquid nitrogen when simulation.}
    \label{Component}	
\end{figure}
\begin{figure}
    \centering
    \includegraphics[width=\columnwidth]{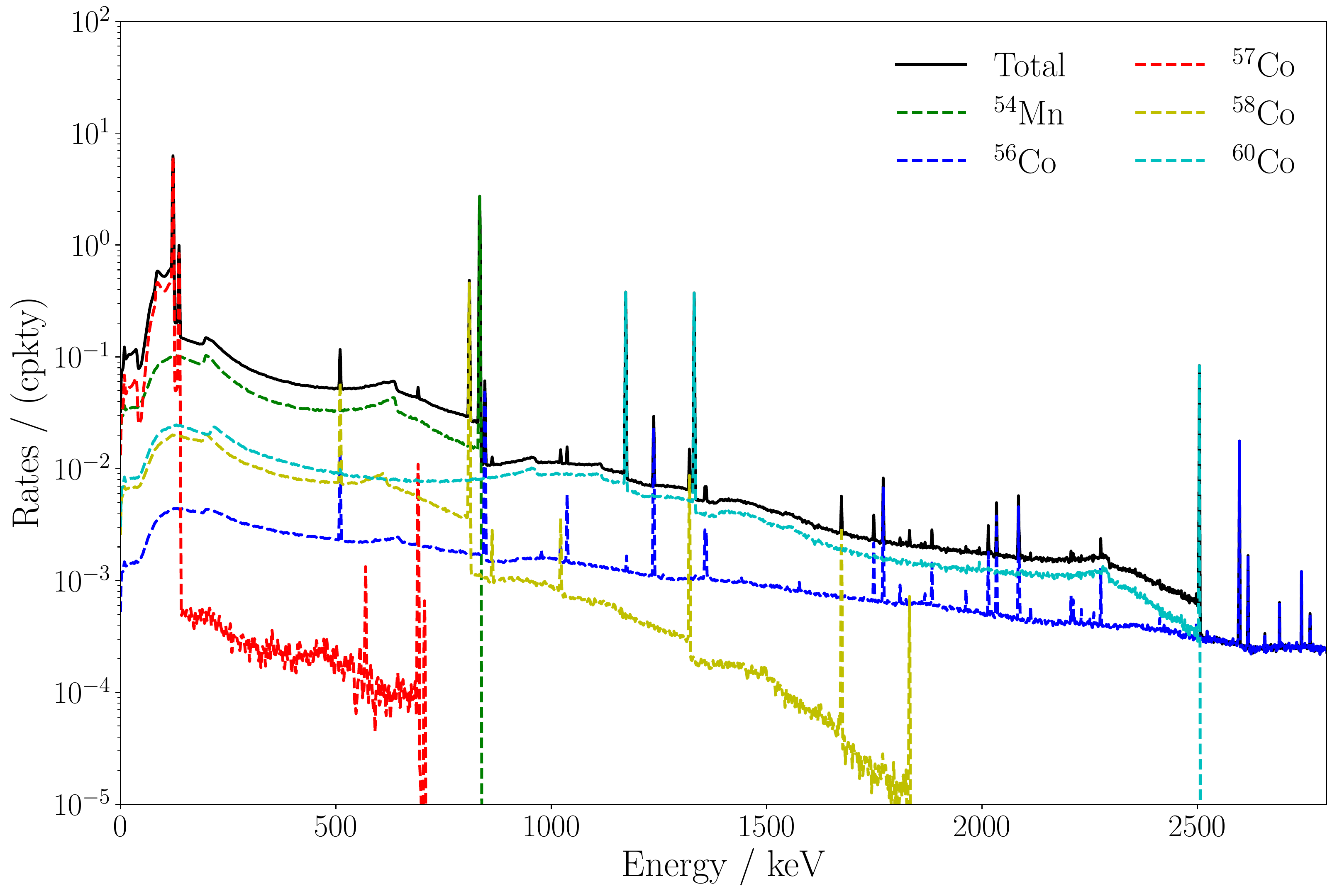}
    \caption{The simulated cosmogenic background spectra of 0-2.8MeV in copper components of germanium detectors for CDEX-100 experiment.The energy resolution is determined by the germanium detector used in CDEX-1B \cite{Yang2019}.}
    \label{SimulationSpec}
\end{figure}

For simplicity, we assume that the copper components follow the same exposure history of the detectors. 
In the expected exposure history, the manufecture time, transportation time and underground storage time are supposed to be 3 months, half a month and one year. 
After 3 month fabrication, the detectors are transported with a shielding, in which the cosmogenic activations can be reduced by a factor of 10 \cite{Ma2018,Abgrall2014}, to the underground laboratory in half a month. 
The altitude of manufacture factory is selected to be at sea level while that of the whole transportation routine is selected to be 1500m, which is the average altitude of Liangshan P
efecture where CJPL is located. 
Then, the cupreous components will be stored in CJPL for preparation work with the decay of cosmogenic nuclides, so-called cooling time, lasting about 1 year. 
Importing the measured production rates of the cosmogenic activations, the background contributions from cupreous components were estimated with a Geant4-based Monte Carlo framework called Simulation and Analysis of Germanium Experiment (SAGE) developed for the CDEX-100 experiment \cite{She2021}.
In the simulation, these cosmogenic radionuclides are distributed uniformly across all cupreous components as shown in Figure~\ref{Component}. 
The energy spectra of the different cosmogenic nuclides and the total background contributions are simulated as shown in Figure~\ref{SimulationSpec}.
\par
Under above assumptions, the simulated background rates in the low energy range (2-4 keV) and around 2039 keV (energy of neutrinoless double beta decay of ${^{76}}$Ge) are ${7.17\times10^{-2}}$ counts per keV per tonne per year (cpkty) and ${1.82\times10^{-3}}$ cpkty, respectively. 
The contribution of ${^{57}}$Co dominates the background contributions lower than 122 keV and ${^{54}}$Mn is dominant when energy is lower than 830 keV.
${^{60}}$Co is prevailing in the higher energy region while there only exists the background from ${^{56}}$Co over 2.5 MeV. 
Among these cosmogenic nuclides, only ${^{60}}$Co has a half-life over one year and it is hard to be significantly suppressed by underground cooling rather than other cosmogenic nuclides do. 
The feasible method to mitigate the contribution from ${^{60}}$Co is to reduce the exposure time above ground or electroform the copper underground. 
\par

\section{Summary} 
We investigate the cosmogenic activation in OFHC copper bricks at the altitude of 2469.4m. 
The specific activities of cosmogenic radionuclides in the exposed copper bricks are measured with a low background germanium gamma-ray spectrometer.
The production rates of several cosmogenic radionuclides are calculated and compared with previous studies. 
The production rates at sea level in units of nuclei/kg/day are ${18.6 \pm 2.0}$ for ${^{54}}$Mn, ${9.9 \pm 1.3}$ for ${^{56}}$Co, ${48.3 \pm 5.5}$ for ${^{57}}$Co, ${51.8 \pm 2.5}$ for ${^{58}}$Co and ${39.7 \pm 5.7}$ for ${^{60}}$Co, respectively. 
For comparison, we simulate their cosmogenic activation with Geant4, showing that the agreement between measurement and simulation is generally satisfactory. 
The discrepancies with other published results are also presented. 
\par
Based on the results of the production rates of cosmogenic radionuclides in copper, the cosmogenic background from copper components of detectors is simulated for CDEX-100 experiment.
The total background rates from cosmogenic nuclides are ${7.2\times10^{-2}}$ cpkty and ${1.8\times10^{-3}}$ cpkty for the 2-4 keV and around 2039 keV energy regions, respectively. 
The cosmogenic background is dominated by ${^{57}}$Co, ${^{54}}$Mn and ${^{60}}$Co after one-year cooling underground. 
To mitigate the cosmogenic activation above ground, it is necessary to control the total exposure time including manufacture and transportation with an appropriate shield. 
Moreover, underground electroforming of copper is another option to dramatically reduce the cosmogenic background from copper components of germanium detectors.
\par

\begin{acknowledgements}
This work was supported by the National Key Research and Development Program of China (No. 2017YFA0402201), National Natural Science Foundation of China (No.11675088, 11725522, U1865205), and Tsinghua University Initiative Scientific Research Program. Thanks to colleagues of CJPL for their help in copper sample measurements.
\end{acknowledgements}
\textbf{Conflicts of interest}: The authors have no relevant financial or non-financial interests to disclose.

\bibliographystyle{unsrt}
\bibliography{reference.bib}

\end{document}